\documentclass[prl,aps,epsfig,superscriptaddress,showpacs,twocolumns]{revtex4}
\usepackage{epsfig,amsmath}
\begin{document}
\author{Jing-Ling Chen}
\email{chenjl@nankai.edu.cn}
\affiliation{Theoretical Physics
Division, Chern Institute of Mathematics, Nankai University, Tianjin
300071, P.R.China} \affiliation{Department of Physics, National
University of Singapore, 2 Science Drive 3, Singapore 117542}
\author{Chunfeng Wu}
\affiliation{Department of Physics, National University of
Singapore, 2 Science Drive 3, Singapore 117542}
\author{L.C. Kwek}
\affiliation{Department of Physics, National University of
Singapore, 2 Science Drive 3, Singapore 117542}
\affiliation{Nanyang
Technological University, National Institute of Education, 1,
Nanyang Walk, Singapore 637616}
\author{C.H. Oh}
\email{phyohch@nus.edu.sg}
\affiliation{Department of Physics,
National University of Singapore, 2 Science Drive 3, Singapore
117542}
\author{Mo-Lin Ge}
\affiliation{Theoretical Physics Division, Chern Institute of
Mathematics, Nankai University, Tianjin 300071, P.R.China}

\title{Violating Bell Inequalities Maximally for Two $d$-Dimensional Systems}
\date{\today}

\begin{abstract}
We investigate the maximal violation of Bell inequalities for two
$d$-dimensional systems by using the method of Bell operator. The
maximal violation corresponds to the maximal eigenvalue of the
Bell operator matrix. The eigenvectors  corresponding to these
eigenvalues are described by asymmetric entangled states. We
estimate the maximum value of the eigenvalue for large dimension.
A family of elegant entangled states $|\Psi\rangle_{\rm app}$ that
violate Bell inequality more strongly than the maximally entangled
state but are somewhat close to these eigenvectors is presented.
These approximate states can potentially be useful for quantum
cryptography as well as many other important fields of quantum
information.
\end{abstract}

 \maketitle

The famous Clauser-Horne-Shimony-Holt (CHSH) inequality
\cite{Bell,CHSH} for two entangled spin-1/2 particles has always
provided an excellent test-bed for experimental verification of
quantum mechanics against the predictions of local realism
\cite{exp}. It is well-known that all pure entangled states of 2
dimension violate the CHSH inequality: the maximum quantum violation
of $2\sqrt{2}$ being often called the Tsirelson's bound
\cite{Gisin}. In 2000, contrary to previous study, Kaszlikowski
\emph{ et al} showed numerically, based on linear optimization
techniques, that the violations of local realism increase with
dimensions for two maximally entangled $d$-dimensional systems
(qudits) ($3\leq d\leq 9$) \cite{Dago00}. A year later, Durt \emph{
et al} extended the analysis to $d=16$ under special experimental
settings\cite{Durt01}. In the same year, an analytical proof was
constructed for two maximally entangled 3-dimensional systems
(qutrits) \cite{JLC1}. In 2002, two research teams independently
developed Bell inequalities for high-dimensional systems: the first
one is a Clauser-Horne type (probability) inequality for two qutrits
\cite{JLC2}; and the second one is a CHSH type (correlation)
inequality to two arbitrary $d$-dimensional systems \cite{CGLMP},
now known as the Collins-Gisin-Linden-Masser-Popoescu (CGLMP)
inequalities. The tightness of the CGLMP inequality was demonstrated
in Ref. \cite{LM}. The maximally entangled state of two-qudit reads
$|\Psi\rangle_{\rm
mes}=\frac{1}{\sqrt{d}}\sum_{j=0}^{d-1}|jj\rangle
$ where $|j\rangle$ is the orthonormal base in each subsystem.
Collins \emph{ et al} restricted their investigation to
$|\Psi\rangle_{\rm mes}$, because on one hand $|\Psi\rangle_{\rm
mes}$ is a simple state, on the other hand there is a very
natural viewpiont that maximally entangled states would maximally
violate the Bell inequality, just as the CHSH inequality has
worked for two-qubit. The results of \cite{CGLMP} was numerically
consistent with Ref. \cite{Dago00}. For $|\Psi\rangle_{\rm mes}$,
the Tsirelson's bound or Bell expression $I_d$ is given by
\cite{CGLMP}
\begin{eqnarray}
I_d(|\Psi\rangle_{\rm mes})  &=&
4d\sum_{k=0}^{\ell}(1-\frac{2k}{d-1})(\frac{1}{2d^3\sin
^2[\pi(k+\frac{1}{4})/d]}\nonumber \\
& & \mbox{\hspace{1 cm}} -\frac{1}{2d^3\sin
^2[\pi(-k-1+\frac{1}{4})/d]}),
 \label{max}
\end{eqnarray}
showing that the maximum quantum violation increases with the
dimension $d$ (here $\ell=[d/2-1]$ represents the integer part of
$(d/2-1)$. ). When $d \rightarrow \infty$, it is interesting to
note that
\begin{eqnarray}
 \lim_{d \rightarrow \infty} I_d(|\Psi\rangle_{\rm mes}) & = &
\frac{2}{\pi^2}
\sum_{k=0}^{\infty}[\frac{1}{(k+1/4)^2}-\frac{1}{(k+3/4)^2}]
\nonumber \\ & \simeq & 2.96981,
 \label{limit}
\end{eqnarray}
a number related to the Catalan's constant.

Contrary to prevalent belief, Ac\'{i}n \emph{ et al} studied the
quantum nonlocality of two-qutrit as well as two $d$-dimensional
systems up to $d=8$ and discovered another unexpected result: there
existed non-maximally entangled states that lead to greater
violation of the CGLMP inequalities compared with maximally
entangled states\cite{Acin}. This surprising result still lacks of a
good intuitive explanation. From the TABLE I of Ref. \cite{Acin},
one observes that the maximal violation increases with dimension
$d$, and it reaches 3.1013 for $d=8$. This gives rise to a natural
question:`` What are the maximal violations of the CGLMP
inequalities if one increases the dimension $d$, especially when $d$
goes to infinity?" The purpose of this Letter is to investigate the
maximal violations of the CGLMP inequalities for higher dimensional
systems. We also present a family of elegant entangled states
$|\Psi\rangle_{\rm app}$ whose corresponding maximal violations are
approximate to the real ones.

Before computing the quantum violation, let us first estimate the
upper bound under quantum mechanics. Significantly, Ref. \cite{Fu}
has recast the CGLMP inequality into a form that is very similar
to the CHSH inequality:
\begin{equation}
I_d=Q_{11}+Q_{12}-Q_{21}+Q_{22}\leq 2,  \label{ch3}
\end{equation}
where $Q_{ij}$ are the correlation functions defined by
probabilities in the following way
\begin{equation}
Q_{ij}= \frac{1}{S}\sum_{m=0}^{d-1}%
\sum_{n=0}^{d-1}f^{ij}(m,n)P(A_{i}=m,B_{j}=n),  \label{qq}
\end{equation}
where $S=(d-1)/2$ is the spin of the particle for the
$d$-dimensional system, $ f^{ij}(m,n)=S-M(\varepsilon
(i-j)(m+n),d)$; $\varepsilon (x)=1$ and $-1$ for $x\geq 0 $ and
$x<0$ respectively; $M(x,d)=(x\; {\rm mod} d)$ and $0\leq
M(x,d)\leq d-1$. Refs. \cite{CGLMP} and \cite{Fu} have proved
that $I_d\leq 2$ for hidden variable theory. From Eq. (\ref{qq}),
one notes that the extreme values of $Q_{ij}$ are $\pm 1$ for
both local realistic description and quantum mechanics, therefore
it is impossible that the maximum quantum violation of $I_d$ is
larger than 4. Furthermore $I_d$ cannot reach 4, because $Q_{ij}$
are constrained to each other, if three of them are set to be 1,
the fourth must also be 1. Consequently, one can conclude easily
from above analysis  that the maximal quantum violation of $I_d$
is a number between 2 and 4.

To generalize Ac\'{i}n's approach, we first note that it has been
shown that unbiased multiport beam splitter \cite{Bellport} can
be used to test violation of local realism of two maximally
entangled qudits. Unbiased $d$-port beam splitter is a device with
the following property: if a photon enters any of the $d$ single
input ports, its chances of exit are equally split  among the $d$
output ports. In fact one can always build the device with the
distinguishing trait that the elements of its unitary transition
matrix $T$ are solely powers of the root of unity
$\gamma=\exp(i2\pi/d)$, namely
$T_{kl}=\frac{1}{\sqrt{d}}\gamma^{kl}$. In front of $i$-th input
port of the device a phase shifter is placed to change the phase
of the incoming photon by $\phi(i)$. These $d$ phase shifts,
denoted for convenience as a ``vector" of phase shifts
$\hat{\phi}=(\phi(0),\phi(1),...,\phi(d-1))$, are macroscopic
local parameters that can be changed by the observer. Therefore,
unbiased $d$-port beam splitter together with the $d$ phase
shifters perform the unitary transformation $U(\hat{\phi})$ with
the entries $U_{kl}=T_{kl}\exp(i\phi(l))$. Devices (Bell
multiports) endowed with such a matrix were proposed, and readers
who are interested in it can refer to Refs.
\cite{Dago00}\cite{Bellport}. The approach developed in
\cite{Acin} is related to Bell operator. An arbitrary entangled
state of two-qudit reads
\begin{eqnarray}
|\Phi\rangle=\sum_{j,j'=0}^{d-1}\alpha_{jj'}|jj'\rangle.
\end{eqnarray}
The quantum prediction of the joint probability $P(A_a=k,B_b=l)$
when $A_a$ and $B_b$ are measured in the initial state
$|\Phi\rangle$ is given by
\begin{widetext}
\vspace{-0.7cm}
\begin{eqnarray}
P(A_a=k,B_b=l)  & = & {\rm Tr}[(U(\hat{\phi}_a)^{\dagger}\otimes
U(\hat{\varphi}_b)^{\dagger})\; \hat{\Pi_k}\otimes \hat{\Pi_l}\;
(U(\hat{\phi}_a)\otimes U(\hat{\varphi}_b))\;
|\Phi\rangle\langle\Phi|] \nonumber  \\
& = & \frac{1}{d^2}\sum_{j,j',m,m'=0}^{d-1}
\alpha_{jj'}\alpha^{*}_{mm'}
e^{i[\phi_a(j)+\varphi_b(j')+\frac{2\pi}{d}(jk-j'l)-\phi_a(m)-\varphi_b(m')-\frac{2\pi}{d}(mk-m'l)]},
\label{subs}
\end{eqnarray}
\end{widetext}
\noindent where $\hat{\Pi_k}=|k\rangle\langle k|$,
$\hat{\Pi_l}=|l\rangle\langle l|$ are the projectors for systems
$A$ and $B$, respectively. Substituting Eq. (\ref{subs}) into the
CGLMP inequality, one gets the Bell expression for the state
$|\Phi\rangle$:
\begin{widetext}
\begin{eqnarray}
I_d(|\Phi\rangle)=&&\frac{1}{d^2}\sum^{d-1}_{j,j',m,m'=0}\alpha_{jj'}\alpha^{*}_{mm'}
\sum_{l=0}^{d-1}e^{i\frac{2\pi}{d}[(j-m)-(j'-m')]l} \nonumber \\
&&\times
\{e^{i[\phi_1(j)-\phi_1(m)+\varphi_1(j')-\varphi_1(m')]}\sum_{k=0}^{\ell}(1-\frac{2k}{d-1})
(e^{i\frac{2\pi}{d}k(j-m)}-e^{-i\frac{2\pi}{d}(k+1)(j'-m')})+ \nonumber \\
&&e^{i[\phi_1(j)-\phi_1(m)+\varphi_2(j')-\varphi_2(m')]}\sum_{k=0}^{\ell}(1-\frac{2k}{d-1})
(e^{-i\frac{2\pi}{d}k(j'-m')}-e^{i\frac{2\pi}{d}(k+1)(j-m)})+ \nonumber \\
&&e^{i[\phi_2(j)-\phi_2(m)+\varphi_1(j')-\varphi_1(m')]}\sum_{k=0}^{\ell}(1-\frac{2k}{d-1})
(e^{-i\frac{2\pi}{d}(k+1)(j'-m')}-e^{i\frac{2\pi}{d}k(j-m)})+ \nonumber \\
&&e^{i[\phi_2(j)-\phi_2(m)+\varphi_2(j')-\varphi_2(m')]}\sum_{k=0}^{\ell}(1-\frac{2k}{d-1})
(e^{i\frac{2\pi}{d}k(j-m)}-e^{-i\frac{2\pi}{d}(k+1)(j'-m')}) \}.
\end{eqnarray}
\end{widetext}
$I_d(|\Phi\rangle)$ can be expressed as
\begin{eqnarray}
I_d(|\Phi\rangle)={\rm
Tr}(\hat{B}|\Phi\rangle\langle\Phi|)=\langle \Phi| \hat{B}
|\Phi\rangle,
\end{eqnarray}
where $\hat{B}$ is the so-called Bell operator. Starting with the
CGLMP inequality and choosing suitable the experimental settings
\cite{Durt01}\cite{Acin}
\begin{eqnarray}
\phi_1(j)=0, \;\;\;\; \phi_2(j)=\frac{j\pi}{d},\;\;\;\;
\varphi_1(j)=\frac{j\pi}{2d}, \;\;\;\;
\varphi_2(j)=-\frac{j\pi}{2d},
 \label{esetting}
\end{eqnarray}
optimal for maximal violations, we can derive the element of the
Bell operator matrix as
\begin{widetext}
\begin{eqnarray}
B_{mm',jj'}=&&\frac{1}{d^2}\sum_{l=0}^{d-1}e^{i\frac{2\pi}{d}[(j-m)-(j'-m')]l} \nonumber \\
&&\{e^{\frac{i\pi}{2d}(j'-m')}\sum_{k=0}^{\ell}(1-\frac{2k}{d-1})(e^{i\frac{2\pi}{d}k(j-m)}
-e^{-i\frac{2\pi}{d}(k+1)(j'-m')})+ \nonumber \\
&&e^{-\frac{i\pi}{2d}(j'-m')}\sum_{k=0}^{\ell}(1-\frac{2k}{d-1})(e^{-i\frac{2\pi}{d}k(j'-m')}
-e^{i\frac{2\pi}{d}(k+1)(j-m)})+ \nonumber \\
&&e^{\frac{i\pi}{d}(j-m)+\frac{i\pi}{2d}(j'-m')}\sum_{k=0}^{\ell}(1-\frac{2k}{d-1})
(e^{-i\frac{2\pi}{d}(k+1)(j'-m')}-e^{i\frac{2\pi}{d}k(j-m)})+ \nonumber \\
&&e^{\frac{i\pi}{d}(j-m)-\frac{i\pi}{2d}(j'-m')}\sum_{k=0}^{\ell}(1-\frac{2k}{d-1})
(e^{i\frac{2\pi}{d}k(j-m)}-e^{-i\frac{2\pi}{d}(k+1)(j'-m')}) \},
\label{findred}
\end{eqnarray}
\end{widetext}
\noindent where $\hat{B}$, in general, is a $d^2 \times d^2$ matrix.
Ref. \cite{Acin} found that the maximal eigenvalue of matrix
$\hat{B}$ is nothing but the highest quantum prediction of the Bell
expression $I_d(|\Phi\rangle_{\rm eig})$ and the corresponding
eigenvector $|\Phi\rangle_{\rm eig}$ is the state that maximally
violates the Bell inequality. Thus, the problem of computing the
maximal violation of $I_d(|\Phi\rangle)$  reduces to the
determination of the maximal eigenvalue of matrix $\hat{B}$. Due to
$\sum_{l=0}^{d-1}e^{i\frac{2\pi}{d}(p-q)l}=d\delta_{pq}$, where
$\delta_{pq}=1$ when $p=q$ modulo $d$ and $0$ otherwise, the matrix
$\hat{B}$ can be further simplified, i.e., it can be decomposed into
the sum of $d$ decoupled operators that act individually within the
subspaces spanned by the vectors
$\{|00\rangle,|11\rangle,...,|(d-1)(d-1)\rangle\}$,
$\{|01\rangle,|12\rangle,...,|(d-1)0\rangle\}$,$...$,$\{|0
(d-1)\rangle,|10\rangle,...,|(d-1)(d-2)\rangle\}$, respectively. For
the first subspace spanned by the vectors
$\{|00\rangle,|11\rangle,...,|(d-1)(d-1)\rangle\}$, and with the
constraint $j-m=j'-m'$, one can reduce $B_{mm',jj'}$ to
\begin{eqnarray}
 & & B^{red1}_{mj}
\nonumber \\
&=&\frac{8}{d}\sin\biggr[\frac{\pi}{2d}(j-m)\biggr] \nonumber \\
& & \times  \sum_{k=0}^{\ell}(1-\frac{2k}{d-1})
\sin\biggr[\frac{2\pi}{d}(k+\frac{1}{2})(j-m)\biggr]\nonumber \\
&=&\frac{2}{d-1}\;
\frac{1}{\cos[\frac{\pi}{2d}(j-m)]}(1-\delta_{mj}) , \label{bred1}
\end{eqnarray}
where $B^{red1}_{mj}$ is the matrix element located at the $m$-th
row and the $j$-th column of the reduced Bell operator $B^{red1}$.
Let us denote $B_r=B_{m,m+r}$, since $B_{m,j}$ depends only on the
difference $(j-m)$. Therefore, we arrive at the following $d
\times d$ real symmetric matrix for the reduced Bell operator:
\begin{eqnarray}
\hat{B}^{red1}&=&\mbox{\hspace{-1mm}}\left(\begin{array}{ccccccc}
B_0 & B_1 & B_2 & \dots & B_{d-3} & B_{d-2} & B_{d-1} \\
B_1 & B_0 & B_1 & B_2 & \dots & B_{d-3} & B_{d-2} \\
B_2 & B_1 & B_0 & \ddots & \ddots & \vdots & B_{d-3} \\
\vdots & B_2 & \ddots & \ddots & \ddots & B_2 & \vdots \\
B_{d-3} & \vdots & \ddots & \ddots & B_0 & B_1 & B_2 \\
B_{d-2} & B_{d-3} & \dots & B_2 & B_1 & B_0 & B_1 \\
B_{d-1} & B_{d-2} & B_{d-3} & \dots & B_2 & B_1 & B_0
\end{array}     \right). \nonumber \\
\end{eqnarray}
One may read from Eq.(\ref{bred1}) that $B_r$ are all non-negative
numbers, $B_0\equiv 0$, and interestingly $\lim_{d \rightarrow
\infty} B_{d-1}=4/\pi$, $\lim_{d \rightarrow \infty}
B_{d-j}/B_{d-1}=1/j$, for $j<< d$. The Bell operator
$\hat{B}^{red1}$ acts on a matrix column $(a_0, a_1, \cdots,
a_{d-1})^T$, which corresponds to the state
$|\Psi\rangle=\sum_{j=0}^{d-1}a_{j}|jj\rangle$ in the first
subspace. Actually, an arbitrary state
$|\Phi\rangle=\sum_{j,j'=0}^{d-1}\alpha_{jj'}|jj'\rangle$ can always
be transformed into its Schmidt decomposition form
$|\Psi\rangle=\sum_{j=0}^{d-1}a_{j}|jj\rangle$ through local unitary
transformations, thus it is sufficient to study the maximal
violation problem in the first subspace. For a general state
$|\Psi\rangle$, one can obtain the Bell expression as

\begin{eqnarray}
I_d(|\Psi\rangle)& = &{\rm
Tr}(\hat{B}^{red1}|\Psi\rangle\langle\Psi|)
=\sum_{m=0}^{d-1}\sum_{j=0}^{d-1}B_{mj}a_m a_j \nonumber \\
& = &\sum_{r=1}^{d-1} B_r\biggr(2 \;\sum_{m=0}^{d-1-r}a_m
a_{m+r}\biggr).
\end{eqnarray}

For instance, for the maximally entangled state $|\Psi\rangle_{\rm
mes}$ with $a_j =1/\sqrt{d}$, the summation
$(\sum_{m,j=0}^{d-1}B_{mj})/d=(\sum_{r=1}^{d-1} 2(d-r)B_r)/d$
recovers the results of Eq. (\ref{max}).

As the analysis shown in \cite{Acin}, the maximal violation of the
CGLMP inequality with the experimental settings (\ref{esetting})
corresponds to the maximal eigenvalue of $\hat{B}^{red1}$, and
indeed its corresponding eigenvector is a nonmaximally entangled
state of two-qudit. For instance, for $d=3$, one has
$B_0=0$,$B_1=2\sqrt{3}/3$,$B_2=2$, the eigenvector
$|\Psi\rangle_{\rm
eig}^{d=3}=\sqrt{\frac{2}{11-\sqrt{33}}}(|00\rangle+\frac{\sqrt{11}-\sqrt{3}}{2}|11\rangle+|22\rangle)$
corresponds to a maximal violation $I_{d=3}(|\Psi\rangle_{\rm
eig})=1+\sqrt{11/3}\simeq 2.9149$, which is larger than
$I_{d=3}(|\Psi\rangle_{\rm mes})\simeq 2.8729$; for $d=4$, one has
$B_0=0$,$B_1=2\sqrt{4-2\sqrt{2}}/3$,$B_2=2\sqrt{2}/3$,
$B_1=2\sqrt{4+2\sqrt{2}}/3$,the eigenvector $|\Psi\rangle_{\rm
eig}^{d=4}=\frac{1}{\sqrt{2+2a^2}}(|00\rangle+a|11\rangle+a|22\rangle+|33\rangle)$,
with
$a=(\sqrt{2+\sqrt{2}}+\sqrt{8-3\sqrt{2}+4\sqrt{2-\sqrt{2}}}-\sqrt{4+2\sqrt{2}})/(\sqrt{2}+\sqrt{4-2\sqrt{2}})\simeq
0.73937$, corresponds to a maximal violation
$I_{d=4}(|\Psi\rangle_{\rm
eig})=\frac{2}{3}\sqrt{2+\sqrt{2}}+\frac{2}{3}\sqrt{8-3\sqrt{2}+4\sqrt{2-\sqrt{2}}}
\simeq 2.9727$, which is larger than $I_{d=4}(|\Psi\rangle_{\rm
mes})\simeq 2.8962$.

By diagonalizing exactly the matrix $\hat{B}^{red1}$, we have
obtained the real maximal violations $I_{d}(|\Psi\rangle_{\rm eig})$
for two entangled qudits. The highest dimension that we have
calculated is $d=8000$
. In Fig. 1, one may observe that $I_{d}(|\Psi\rangle_{\rm eig})$
increases with dimension $d$. Based on the data of
$I_{d}(|\Psi\rangle_{\rm eig})$ from $d=2$ to $d=8000$, one has an
empirical formula fitting $I_{d}(|\Psi\rangle_{\rm eig})$
numerically to the dimension $d$:
\begin{eqnarray}
I_{d}^{\rm rough}(|\Psi\rangle_{\rm eig})&\simeq&
3.9132 - 1.2891 x^{-0.2226}
\end{eqnarray}
from which one can see that $I_{d}^{\rm rough}(|\Psi\rangle_{\rm
eig})\simeq 3.9132$ is a coarse-grained limit of the maximal
violation for the CGLMP inequality when $d$ tends to infinity.

\begin{figure}
\begin{center}
\epsfig{figure=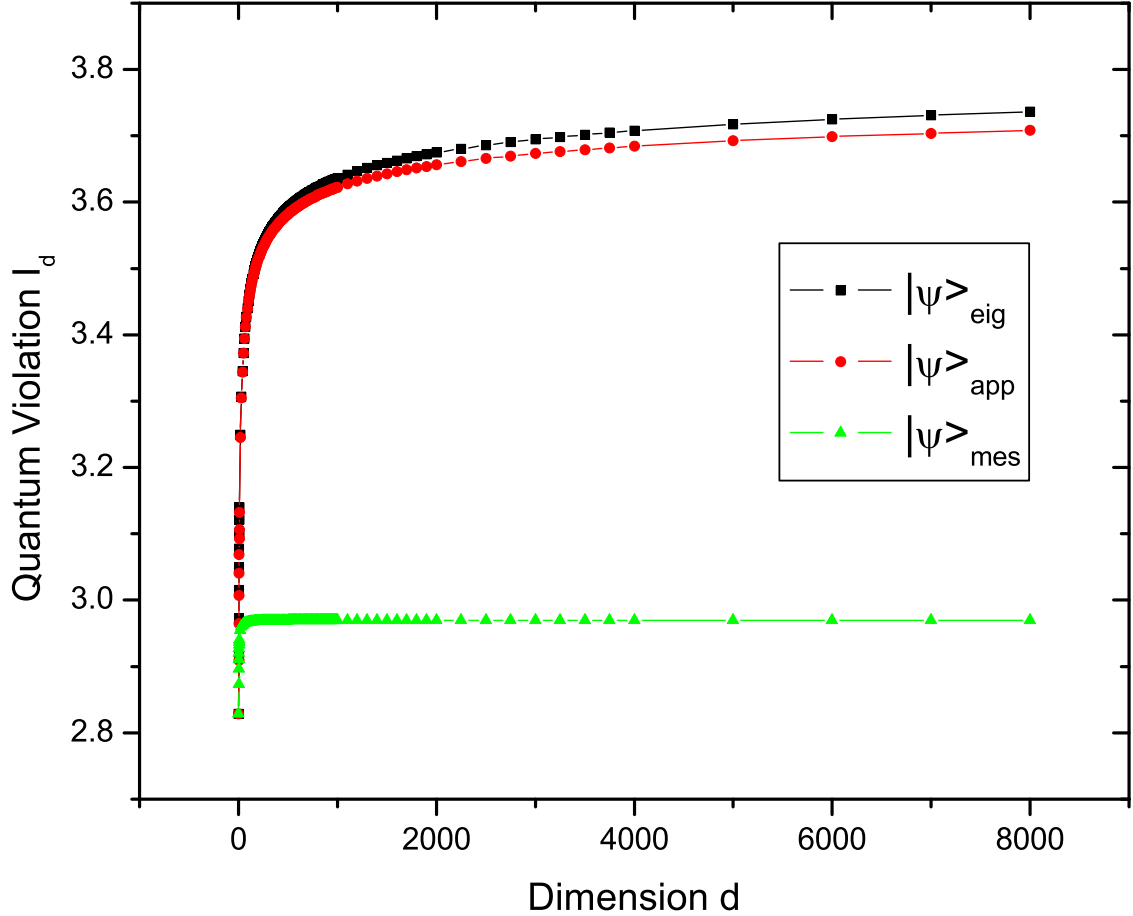,width=\columnwidth}
\end{center}
\caption{Variations of $I_d(|\Psi\rangle)$ with increasing
dimension $d$ ($2 \leq d \leq 8000$).} \label{2qunitmaximalfig}
\end{figure}
Analysis of the eigenvectors $|\Psi\rangle_{\rm eig}$ shows that
these eigenvectors numerically satisfy some general properties:
for instance, $|\Psi\rangle_{\rm eig}=\sum_{j=0}^{d-1} a_j^{\rm
eig}|jj\rangle$ with maximal eigenvalue has the following
symmetric properties for the coefficients: $a_j=a_{d-1-j}$; and
$a_0:a_1:a_2:a_3:\cdots \simeq
1:\frac{1}{\sqrt{2}}:\frac{1}{\sqrt{3}}:\frac{1}{\sqrt{4}}:\cdots$
for large $d$. Thus, we may approximate a family of elegant
entangled states
\begin{eqnarray}
|\Psi\rangle_{\rm app}& = & \sum_{j=0}^{d-1} a_j^{\rm
app}|jj\rangle, \;\; a_j^{\rm app}=\frac{1}{\sqrt{\cal N}}\;
\frac{1}{\sqrt{(j+1)(d-j)}}, \;\; \nonumber \\ & & {\cal N}=
\sum_{j=0}^{d-1}\frac{1}{(j+1)(d-j)}
 \label{2qunitapp}
\end{eqnarray}
whose corresponding Bell expressions $I_d(|\Psi\rangle_{\rm app})$
are closed to the actual ones $I_d(|\Psi\rangle_{\rm eig})$. For
example, for $d=8000$, the error rate between
$I_d(|\Psi\rangle_{\rm eig})$ and $I_d(|\Psi\rangle_{\rm app})$
is only about $0.745 \%$. We have also listed
$I_d(|\Psi\rangle_{\rm app})$ and $I_d(|\Psi\rangle_{\rm mes})$
in Table I and drawn the corresponding curves in Fig. 1. One may
observe that for $d=50000$, $I_d(|\Psi\rangle_{\rm mes})\simeq
2.96981$ has almost reached the limit as shown in Eq.
(\ref{limit}); and for $d=600000$, $I_d(|\Psi\rangle_{\rm app})$
has exceeded 3.80.

In summary, we have investigated the maximal violation of the
CGLMP inequalities for two entangled qudits. The maximal
violation occurs at the non-maximally entangled state, which is
the eigenvector of Bell operator with the maximal eigenvalue. The
maximal violations increase with growing dimension $d$, coming
close to 4 when $d$ approaches to infinity. Experimental test has
been performed to verify the CGLMP inequalities for the first few
high-dimensional systems, and indeed there exist non-maximally
entangled states  that violate these inequalities more strongly
than the maximal entangled ones \cite{Vaziri}.  Bell inequalities
are applicable to quantum cryptography and quantum communication
complexity \cite{Kwek}, previous researches are mostly based on
maximally entangled states. For stronger violation, it may be
useful to generate the approximate state in this paper.
Nevertheless, it may be significant and interesting to apply the
elegant symmetric entangled states $|\Psi\rangle_{\rm app}$ to
quantum cryptography as well as other important fields of quantum
information.

This work is supported by NUS academic research Grant No. WBS:
R-144-000-123-112.


\begin{table}
\begin{tabular}{|c|c|c|c|c|c|c|c|c|c|c|c|c|c|c|}
     \hline\hline
$d$   &$2$ &$3$   &$4$      &$5$      &$6$      &$7$ &$8$
&$9$      &$10$     &$20$     &$30$     &$40$
&$50$     &$60$\\
     \hline
$I_d(|\Psi\rangle_{\rm eig})$ &$2.82843$  &$2.9149$
&$2.9727$ &$3.0157$ &$3.0497$ &$3.0777$ &$3.1013$ &$3.1217$ &$3.1396$ &$3.2492$ &$3.3068$ &$3.3449$ &$3.3728$ &$3.3946$\\
     \hline
$I_d(|\Psi\rangle_{\rm app})$ &$2.82843$  &$2.90909$ &$2.96466$
&$3.00689$ &$3.04078$ &$3.06895$ &$3.09296$ &$3.10639$ &$3.13219$
&$3.24551$ &$3.30496$ &$3.34389$
&$3.37222$ &$3.39417$\\
     \hline
$I_d(|\Psi\rangle_{\rm mes})$ &$2.82843$  &$2.87293$ &$2.89624$
&$2.91054$ &$2.92020$
&$2.92716$ &$2.93241$ &$2.93651$ &$2.93980$ &$2.95472$ &$2.95974$ &$2.96225$ &$2.96376$ &$2.96477$\\
      \hline\hline
$d$          &$70$     &$80$     &$90$     &$100$    &$150$
&$200$    &$250$    &$300$
&$350$    &$400$    &$450$    &$500$    &$550$    &$600$\\
     \hline
$I_d(|\Psi\rangle_{\rm eig})$
 &$3.4124$ &$3.4273$ &$3.4400$   &$3.4511$ &$3.4914$ &$3.5178$ &$3.5370$  &$3.552$
 &$3.5641$ &$3.5743$ &$3.583$  &$3.5906$ &$3.5973$ &$3.6033$ \\
     \hline
$I_d(|\Psi\rangle_{\rm app})$
 &$3.41192$ &$3.4267$ &$3.4393$   &$3.45022$ &$3.48933$ &$3.51446$ &$3.53256$  &$3.54651$
&$3.55775$ &$3.5671$ &$3.57505$  &$3.58194$ &$3.588$ &$3.59339$ \\
     \hline
$I_d(|\Psi\rangle_{\rm mes})$ &$2.96549$  &$2.96603$ &$2.96645$
&$2.96678$ &$2.96779$
&$2.96830$ &$2.96860$ &$2.96880$ &$2.96895$ &$2.96906$ &$2.96914$ &$2.96921$ &$2.96926$ &$2.96931$\\
       \hline\hline
$d$         &$650$    &$700$    &$750$    &$800$
&$850$    &$900$    &$950$    &$1000$   &$1500$   &$2000$   &$2500$   &$3000$   &$3500$   &$4000$\\
     \hline
$I_d(|\Psi\rangle_{\rm eig})$
 &$3.6087$ &$3.6136$ &$3.6181$ &$3.6222$
 &$3.6260$  &$3.6296$ &$3.6329$ &$3.6360$  &$3.6594$ &$3.6747$ &$3.6859$ &$3.6946$ &$3.7017$ &$3.7077$\\
     \hline
$I_d(|\Psi\rangle_{\rm app})$
 &$3.59824$ &$3.60263$ &$3.60664$ &$3.61032$
 &$3.61372$  &$3.61687$ &$3.61981$ &$3.62256$  &$3.64299$ &$3.65623$ &$3.66584$ &$3.67330$ &$3.67936$ &$3.68442$\\
      \hline
$I_d(|\Psi\rangle_{\rm mes})$ &$2.96935$  &$2.96938$ &$2.96941$
&$2.96944$ &$2.96946$
&$2.96948$ &$2.96950$ &$2.96951$ &$2.96961$ &$2.96966$ &$2.96969$ &$2.96971$ &$2.96973$ &$2.96974$\\
      \hline\hline
$d$         &$5000$    &$6000$  &$7000$   &$8000$ &$50000$
 &$70000$
&$80000$   &$90000$   &$100000$   &$200000$   &$300000$   &$400000$   &$500000$ &$600000$\\
     \hline
$I_d(|\Psi\rangle_{\rm eig})$
 &$3.7174$ &$3.7250$ &$3.7311$ &$3.7362$
 &$$  &$$ &$$ &$$  &$$ &$$ &$$ &$$ &$$ &$$\\
     \hline
$I_d(|\Psi\rangle_{\rm app})$
   &$3.69253$    &$3.69884$   &$3.70398$ &$3.70829$
&$3.75659$        &$3.76372$  &$3.76644$    &$3.76878$   &$3.77083$   &$3.78345$   &$3.79019$   &$3.79472$   &$3.79810$ &$3.80080$ \\
      \hline
$I_d(|\Psi\rangle_{\rm mes})$
  &$2.96975$    &$2.96976$  &$2.96977$  &$2.96978$
&$2.96981$    &$2.96981$    &$2.96981$    &$2.96981$   &$2.96981$
&$2.96981$   &$2.96981$   &$2.96981$   &$2.96981$
   &$2.96981$\\
      \hline\hline
          \end{tabular}
          \caption{[Optional Table (for referee)] Bell expressions $I_d(|\Psi\rangle_{\rm eig})$,
     $I_d(|\Psi\rangle_{\rm app})$ and $I_d(|\Psi\rangle_{\rm mes})$.}\label{tab1}
          \end{table}

\end{document}